\documentclass[
nopreprint,
]{jasatex}

\usepackage{graphicx}
\usepackage{dcolumn}
\usepackage{amsmath,amsfonts}
\usepackage{bm}

\begin{document}


\title[Monitoring weak changes in concrete]{Monitoring stress related velocity variation in concrete with a $2.10^{-5}$ relative resolution using diffuse ultrasound.}

\author{Eric LAROSE}
\affiliation{Laboratoire de G\'eophysique Interne et Tectonophysique, CNRS \& Universit\'e J. Fourier, BP53, 38041 Grenoble, France. Email: eric.larose@ujf-grenoble.fr}

\author{Stephen HALL}%
\affiliation{Laboratoire 3S-R, CNRS \& Grenoble Universities, Grenoble, France}

\date{\today}

\begin{abstract}
Ultrasonic waves propagating in solids have stress-dependent velocities. The relation between stress (or strain) and velocity forms the basis of non-linear acoustics. In homogeneous solids, conventional time-of-flight techniques have measured this dependence with spectacular precision. In heterogeneous media like concrete, the direct (ballistic) wave around 500~kHz is strongly attenuated and conventional techniques are less efficient. In this manuscript, the effect of weak stress changes on the late arrivals constituting the acoustic diffuse coda is tracked. A resolution of $2.10^{-5}$ in relative velocity change is attained which corresponds to a sensitivity to stress change of better than 50 kPa. Therefore the technique described here provides an original way to measure the non-linear parameter with stress variations on the order of tens of kPa.  

\end{abstract}

\maketitle

\begin{figure}
		\includegraphics[width=8cm]{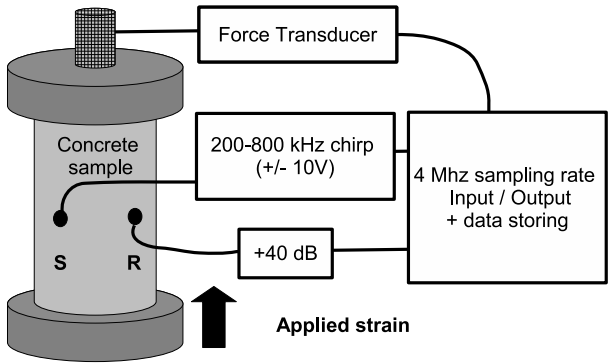}
	\caption{Schematic view of the experimental setup.}
	\label{fig1}
\end{figure}

Concrete is a complex heterogeneous mix of various ingredients of different sizes ranging from centimetric to millimetric gravel and sand, to the micrometric porosity of the cement paste. Over the last thirty years or so, various sonic and ultrasonic techniques have been developed to image, characterize or control the material\cite{popovics1994,abraham2000,mccann2001,payan2007}. Applications mostly work in the low frequency regime (a few Hz up to 50 kHz), where the wavelength is greater than the sizes of heterogeneities. In this regime, direct waves do not suffer too much from scattering by the heterogeneities and wave arrivals can be simply related to ballistic (direct or simply reflected) propagation paths. Nevertheless, because the wavelengths are much greater than the microstructure of the concrete, the fine details like micro-damage or millimetric cracks are hardly visible. In the high frequency regime (above 50 kHz), the elastic wavefield is sensitive to these small details, but the ultrasonic wavelength is also of the order of the aggregates' size. This results in a strong attenuation of the direct (ballistic) wave \cite{landis1995,philippidis2005}, and the appearance of a long lasting coda made of waves that have been multiply scattered by the internal microstructure\cite{anugonda2001,becker2003}. The multiple scattering regime is often associated with a catastrophic loss of information, such that most conventional imaging or monitoring techniques fail. Nevertheless, as diffuse waves travel much longer paths than direct or simply reflected ones, they are much more sensitive to weak perturbation of the medium. This idea has been exploited in geophysics to quantify weak changes in the velocity of the earth crust\cite{poupinet1984}: by comparing two seismic coda obtained from fixed source and receiver at two different times, it is possible to monitor weak velocity variations in the medium. This technique is known as the \textit{seismic doublet} technique or (more recently) \textit{coda wave interferometry}\cite{snieder2002}.

Various phenomena can cause acoustic velocity variations in concrete, including chemical reactions, and changes in water content, temperature or stress. In this work we focus on the latter effect, i.e., the relation between stress/strain and acoustic velocity, which is known as the acousto-elastic effect\cite{murnaghan1951,egle1976,mi2005}. 
 Under vertical uniaxial stress $\Delta \sigma$ the ultrasonic velocities in a sample will evolve as (to first order):
\begin{equation}
V_{ij}=V_{ij}^0  +\frac{\partial V_{ij}}{\partial \sigma}\Delta \sigma +o(\epsilon),
\end{equation}
where $ij$ stands for the wave velocity in direction $i$ for a particle motion in $j$. The dimensionless non-linear parameter $\beta$ is defined as:
\begin{equation}
\beta_{ij}=-\frac{E}{V^0_{ij}}\frac{\partial V_{ij}}{\partial \sigma}
\end{equation}
and depends on the second (Lam\'e) and third (Murnaghan) order elastic constants. $E\approx 3.9.10^{10}Pa$ is the Young's modulus. 
There has been much less attention on the $\sigma$-dependence of $V$ in heterogeneous media, such as concrete\cite{wu1998}, than in homogeneous media, and most results have considered variations under significant applied stress. To our knowledge, the use of diffuse ultrasound to evaluate the effect of stress loading on velocity variations has only been reported by Gr\^et et al.\cite{gret2006} and Payan \cite{payan:phd}.
  In these works the multiple time-window \textit{doublet} technique was used to infer relative velocity variation with a relative resolution of $10^{-3}$ to $10^{-4}$ for stress variation of the order of 1~MPa. Here, we propose to monitor weaker relative velocity variations in concrete down to $2.10^{-5}$ under weaker load variations (50~kPa). Such precision requires high frequency data (here 500~kHz diffuse ultrasonic data) and a more sensitive data processing technique than the \textit{doublet} technique.\\

\begin{figure}
		\includegraphics[width=8cm]{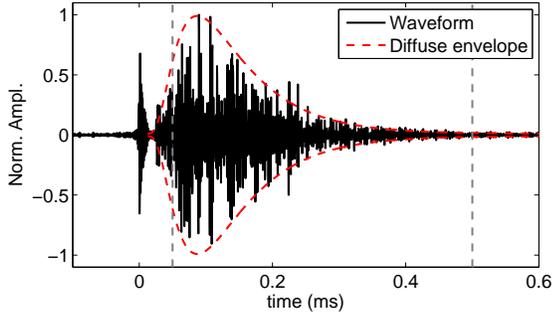}
	\caption{Typical waveform $h[t]$ collected through the concrete sample.  The processed time-window marked by two vertical dashed lines extends from 50~$\mu s$ to 500~$\mu s$.}
	\label{fig2}
\end{figure}

The experimental data presented here were acquired on a concrete sample of 16~cm in diameter, and 28.5~cm in height. The concrete was prepared with 17\% (in weight) of cement, 31\% of fine sand, 43\% of gravel (size ranging from 2 to 10~mm) and 9\% of water. Two ultrasonic transducers are used in the experiment, one as a source and one as a receiver (labeled S and R, respectively, in Fig.~\ref{fig1}). The lateral size of the active part of the transducers (0.7~mm) is much smaller than the wavelength, which makes them very sensitive to multiply scattered waves. The transducers are glued onto the sample using a hot chemical glue (phenyl-salicylic acid) that solidifies with cooling (below 43$^\circ$C). The ultrasonic experiment was performed several months after casting the sample.

During the ultrasonic measurements the concrete specimen was subjected to uniaxial loading. A preliminary load was applied (5 MPa) at which point the reference ultrasonic data $r_0(t)$ were acquired. The applied uniaxial load was then increased to 5.5~MPa in ten step of 50~kPa with ultrasonic acquisition at each stress increments, $k$, to provide $r_k(t)$. Each ultrasonic acquisition involved a source excited with a 10V chirp $s(t)$ of frequencies ranging linearly from 200~kHz to 800~kHz and of duration 10~s. The received signal $r_k(t)$ was amplified by 40~dB and stored in the computer.  The recorded waveforms were subsequently cross-correlated with $s(t)$ (time-compression) to provide an estimate of the impulse response of the sample $h_k(t)=r_k(t)\times s(t)$, for each increment $k$. The effective duration of the time-compressed signal $h_k(t)$ is about 0.6 ms. $h_0(t)$ is the reference waveform corresponding to $r_0(t)$ correlated with $s(t)$. It is noted that the source and receiver positions were kept fixed for all acquisitions and the excitation waveform $s(t)$ was not changed. Note also that	the velocity change due to strain is one order of magnitude smaller than the one due to stress change, and is neglected here. The time separating two consecutive measurements is 5 minutes, over which no stress change was recorded and the ultrasonic records were reproducible. Creep was observed over much longer time scales (several hours) and was negligible in the present experiment. 

A typical ultrasonic record is plotted in Fig.~\ref{fig2}. This record is composed of: (1) the impulse response  of the concrete sample; (2) a peak at t=0 marking the initial excitation and due to the electronic cross-talk between S and R; (3) background electronic noise (about 1\% relative amplitude). The impulse response consists of a direct wave arriving at 0.023~ms, which is barely visible, and a following coda due to strong multiply scattered waves. The average intensity of the coda, $I(r,t)$, can be described by a diffusion envelope: 
 \begin{equation}
  I(r,t)=\frac{I_0}{8(\pi Dt)^{3/2}}e^{-\frac{r^2}{4Dt}-\gamma t},\label{eq_diffusion}.
  \end{equation}
where $I_0$ is the energy released by the source, $D$ is the diffusion constant, $r$ is the source-receiver distance and $\gamma$ is the dissipation rate. In this experiment around 500~kHz, we find that $\sqrt{I}$ best fits the data for $D=5~mm^2/\mu s$ and $\gamma = 25~ms^{-1}$ .  This will represent a slight under-estimate for $D$ because Eq.~\ref{eq_diffusion} only holds for unbounded media. Nevertheless, $D$ and $\sigma$ values are in agreement with the literature \cite{anugonda2001,becker2003}, and our rough estimation of $D$ confirms that we are in the strong multiple scattering regime.

\begin{figure}
		\includegraphics[width=8cm]{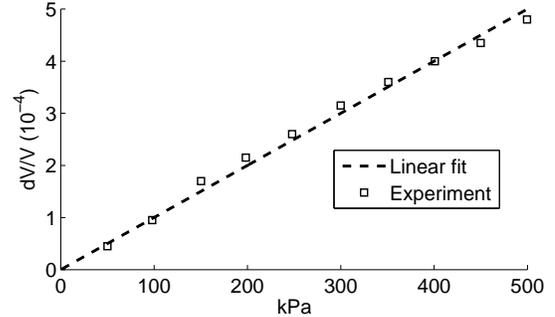}
	\caption{Relative velocity variations $dV/V$ versus stress variation $\Delta\sigma$. }
	\label{fig3}
\end{figure}

For each load $k$, the record $h_k(t)$ is compared to the reference waveform $h_0(t)$ to evaluate the relative velocity change in the concrete sample. Because of strong multiple scattering, the P- and S-wave arrivals mix together and rapidly attain the equipartition regime\cite{weaver1982}. Let $a_{ij}$ be the relative fraction of time that the wave spends in the state $ij$ (polarization in the direction $i$ while propagating in the direction $j$), then the velocity variation is $dV=\left\langle a_{ij}\frac{\partial V_{ij}}{\partial \sigma}\right\rangle \Delta \sigma$. Note that we have $\sum a_{ij}=1$ and that $\left\langle .\right\rangle$ denotes ensemble averaging. If the waves are statistically equipartitioned in all directions, we have $\sum a_{i=j}=9\%$ and $\sum  a_{i\neq j}=91\%$\cite{weaver1982}. In general, $a_{ij}$ has a more complex distribution, but in this work we focus on the average quantity $\left\langle a_{ij}\beta_{ij}\right\rangle$.
For simplicity, we also assume that the waveforms are just stretched in time. This is again a first order approximation that applies to experimental data in the considered time-window. Nevertheless, velocity variations will progressively decorrelate the waveform\cite{lobkis2003}. This weak decorrelation, also called distortion, is not studied here although it contains precious information on the medium and its evolution. In particular, we stress that strong distortion and weak stretching might indicate a change of structure rather than a global velocity change. However here we do not see significant distortion. To summarize, the velocity change $dV$  for load $k$ results in:
\begin{equation}
h_k[t]= h_0\left[t(1+\nu_k)\right]+n(t)\label{h0h}
\end{equation}
with $\nu_k=dV/V$ and $n(t)$ accounting for additional fluctuations, including electronic noise and the small distortion mentioned earlier. Two processing techniques have been proposed in the literature to estimate $dV/V$. The first one, called the seismic \textit{doublet} technique, was developed for geophysical purposes about twenty years ago \cite{poupinet1984}. The idea is to measure a time-shift between two different records in a limited time-window centered at time $t$. By measuring such time-shifts (or delay) $\delta t$ at different times $t$ in the coda, it is possible to evaluate the velocity variation, which is simply the average slope of $\delta t (t)$: $dV/V=-\delta t/t$. Such an approach implicitly assumes that the time-shift is constant within the considered time-window, which is generally not the case. Another idea\cite{lobkis2003,sens-schoenfelder2006} is to interpolate $h_k$ at times $t(1-\nu)$ with various relative velocity changes $\nu$\cite{note1}. Therefore $\nu_k$ is the $\nu$ that maximizes the cross-correlation coefficient:
\begin{equation}
CC_k(\nu)=\frac{\int_{0}^{T}h_k\left[t(1-\nu)\right]h_0[t]dt}{\sqrt{\int_{0}^{T}h_k^2\left[t(1-\nu)\right]dt \int_{0}^{T}h_0^2[t]dt}}.
\end{equation}

Contrary to the \textit{doublet} technique, we do not assume a constant time-shift in the considered time window [0 T]. The comparison between the two waveforms can therefore be performed over the entire record at once, which results in a more stable, and thus more precise, estimation of $dV/V$. The comparison between the \textit{doublet} and the \textit{stretching} technique will be subject to further investigation\cite{hadziioannou2009}. Assuming that $h_0$ and $h_i$ are stationary waveforms and are well described by Eq.~\ref{h0h}, we have a theoretical estimation of $CC$:
\begin{equation}
CC_k(\nu)=A \frac{\int_{\Delta f} \rho(f) sinc  \left(2\pi f \left(\nu-\nu_k\right) T\right)   df}{\int_{\Delta f} \rho(f) df} +B(\nu)
\end{equation}
with $f$ the frequency, $\Delta f$ the bandwidth, $\rho(f)$ the spectral density. The constant $A$ depends on the variance of $h$, noted $\left\langle h^2\right\rangle$ and the variance of the additional fluctuations, noted $\left\langle n^2\right\rangle$,
\begin{equation}
A=\frac{\left\langle h^2\right\rangle}{\left\langle h^2\right\rangle+\left\langle n^2\right\rangle}.
\end{equation}

 $B(\nu)$ is a random process of zero mean  and standard deviation:
\begin{equation}
\sqrt{\left\langle B^2\right\rangle}=\frac{1}{\sqrt{T\Delta f}}\frac{\left\langle n^2\right\rangle+2\sqrt{\left\langle n^2\right\rangle\left\langle h^2\right\rangle}}{\left\langle h^2\right\rangle+\left\langle n^2\right\rangle}.
\end{equation}

If the amplitude of the signal $A$ is much greater than the fluctuations $\sqrt{\left\langle B^2\right\rangle}$ of the cross-correlation coefficient $CC_k$, the maximum of $CC_k$ is obtained for the stretching $\nu=\nu_k$. It is interesting to note that the peak of the $sinc$ function is visible even if the fluctuations (or noise) $n(t)$ are strong. In such a case, increasing the integration time $T$ or the frequency bandwidth $\Delta f$ can compensate for this strong noise. This is a crucial advantage of the present technique compared to the \textit{doublet} technique. The $dV/V$ resulting from our measurements are plotted in Fig.~\ref{fig3} versus $\Delta \sigma$. A linear regression of the data yields $\left\langle \frac{1}{V}\frac{\partial V}{\partial \sigma}\right\rangle=10^{-6}~kPa^{-1}$, which gives an estimate of the non-linear parameter of our sample: $\beta=\left\langle a_{ij}\beta_{ij}\right\rangle\approx -40$. The standard deviation from the linear regression for the relative velocity change $dV/V$ is $2.10^{-5}$, which indicates that the precision on the evaluation of $\beta$ is 5\%. Note that during the whole experiment, the surrounding air temperature fluctuated within less than 0.8$^{\circ}$C. Temperature variations are too weak and thermal diffusion too slow in the sample (compared to the duration of the experiment) to induce noticeable $dV$\cite{snieder2002,lobkis2003,larose2006,leroy2008}. This assertion is confirmed by the fact that the velocity variations $dV$ are totally decorrelated from the air temperature variations. We therefore propose that the variations are due primarily to stress effects.

A potential application of this technique is on-site assessment of the non-linear constant $\beta$ of a concrete structure. Our technique is sensitive enough to provide reliable measurements with additional stress $\Delta \sigma$ of the order of 100 kPa. As the non-linear parameter strongly depends on the micro-damage of the concrete, our technique would form an efficient and non-destructive way to probe the integrity of civil engineering structures. Another application is to evaluate the state of stress of a given structure with a predetermined $\beta$ parameter, and can be extended to geophysical media\cite{yamamura2003,niu2008,brenguier2008b}.


\begin{acknowledgments}
We are thankful to P. Roux, M. Campillo, S. Catheline, R. Weaver for fruitful discussions and J.B. Toni for experimental help. This work was partially funded by UJF-TUNES and RNVOR programs.
\end{acknowledgments}


\begin{thebibliography}{10}
\newcommand{\enquote}[1]{``#1''}
\expandafter\ifx\csname url\endcsname\relax
  \def\url#1{\texttt{#1}}\fi
\expandafter\ifx\csname urlprefix\endcsname\relax\def\urlprefix{URL }\fi
\providecommand{\bibinfo}[2]{#2}
\providecommand{\noopsort}[1]{}
\providecommand{\switchargs}[2]{#2#1}

\bibitem{popovics1994}
\bibinfo{author}{J.~S. Popovics} and \bibinfo{author}{J.~L. Rose},
  \enquote{\bibinfo{title}{A survey of developments in ultrasonic {NDE} of
  concrete}}, \bibinfo{journal}{IEEE Trans. Ultrason., Ferroelec., Freq.
  Contr.} \textbf{\bibinfo{volume}{41}}, \bibinfo{pages}{140--143}
  (\bibinfo{year}{1994}).

\bibitem{abraham2000}
\bibinfo{author}{O.~Abraham}, \bibinfo{author}{C.~Leonard},
  \bibinfo{author}{P.~Cote}, and \bibinfo{author}{B.~Piwakowski},
  \enquote{\bibinfo{title}{Time frequency analysis of impact-echo signals:
  Numerical modeling and experimental validation}}, \bibinfo{journal}{ACI Mat.
  J.} \textbf{\bibinfo{volume}{97}}, \bibinfo{pages}{647--655}
  (\bibinfo{year}{2000}).

\bibitem{mccann2001}
\bibinfo{author}{D.~M. McCann} and \bibinfo{author}{M.~C. Forde},
  \enquote{\bibinfo{title}{Review of {NDT} methods in the assessment of
  concrete and masonry structures}}, \bibinfo{journal}{NDTE Int.}
  \textbf{\bibinfo{volume}{34}}, \bibinfo{pages}{71--84}
  (\bibinfo{year}{2001}).

\bibitem{payan2007}
\bibinfo{author}{C.~Payan}, \bibinfo{author}{V.~Garnier},
  \bibinfo{author}{J.~Moysan}, and \bibinfo{author}{P.~A. Johnson},
  \enquote{\bibinfo{title}{Applying nonlinear resonant ultrasound spectroscopy
  to improving thermal damage assessment in concrete}}, \bibinfo{journal}{JASA
  Express Lett.} \textbf{\bibinfo{volume}{121}}, \bibinfo{pages}{EL125--EL130}
  (\bibinfo{year}{2007}).

\bibitem{landis1995}
\bibinfo{author}{E.~N. Landis} and \bibinfo{author}{S.~P. Shah},
  \enquote{\bibinfo{title}{Frequency-dependent stress wave attenuation in
  cement-based materials}}, \bibinfo{journal}{J. Eng. Mech ASCE}
  \bibinfo{pages}{737--743} (\bibinfo{year}{1995}).

\bibitem{philippidis2005}
\bibinfo{author}{T.~P. Philippidis} and \bibinfo{author}{D.~G. Aggelis},
  \enquote{\bibinfo{title}{Experimental study of wave dispersion and
  attenuation in concrete}}, \bibinfo{journal}{Ultrasonics}
  \textbf{\bibinfo{volume}{43}}, \bibinfo{pages}{584-- 595}
  (\bibinfo{year}{2005}).

\bibitem{anugonda2001}
\bibinfo{author}{P.~Anugonda}, \bibinfo{author}{J.~S. Wiehn}, and
  \bibinfo{author}{J.~A. Turner}, \enquote{\bibinfo{title}{Diffusion of
  ultrasound in concrete}}, \bibinfo{journal}{ultrasonics}
  \textbf{\bibinfo{volume}{39}}, \bibinfo{pages}{429--435}
  (\bibinfo{year}{2001}).

\bibitem{becker2003}
\bibinfo{author}{J.~Becker}, \bibinfo{author}{L.~J. Jacobs}, , and
  \bibinfo{author}{J.~Qu}, \enquote{\bibinfo{title}{Characterization of
  cement-based materials using diffuse ultrasound}}, \bibinfo{journal}{J. Eng.
  Mech.} \textbf{\bibinfo{volume}{129}}, \bibinfo{pages}{1478--1484}
  (\bibinfo{year}{2003}).

\bibitem{poupinet1984}
\bibinfo{author}{G.~Poupinet}, \bibinfo{author}{W.~L. Ellsworth}, and
  \bibinfo{author}{J.~Frechet}, \enquote{\bibinfo{title}{Monitoring velocity
  variations in the crust using earthquake doublets: an application to the
  calaveras fault, california}}, \bibinfo{journal}{J. Geophys. Res.}
  \textbf{\bibinfo{volume}{89}}, \bibinfo{pages}{5719} (\bibinfo{year}{1984}).

\bibitem{snieder2002}
\bibinfo{author}{R.~Snieder}, \bibinfo{author}{A.~Gr\^et},
  \bibinfo{author}{H.~Douma}, and \bibinfo{author}{J.~Scales},
  \enquote{\bibinfo{title}{Coda wave interferometry for estimating nonlinear
  behavior in seismic velocity}}, \bibinfo{journal}{Science}
  \textbf{\bibinfo{volume}{295}}, \bibinfo{pages}{2253} (\bibinfo{year}{2002}).

\bibitem{murnaghan1951}
\bibinfo{author}{F.~D. Murnaghan}, \emph{\bibinfo{title}{Finite Deformation of
  an Elastic Solid.}} (\bibinfo{publisher}{John Wiley, New York}) (\bibinfo{year}{1951}).

\bibitem{egle1976}
\bibinfo{author}{D.~M. Egle} and \bibinfo{author}{D.~E. Bray},
  \enquote{\bibinfo{title}{Measurement of acoustoelastic and 3rd-order
  elastic-constants for rail steel}}, \bibinfo{journal}{J. Acoust. Soc. Am.}
  \textbf{\bibinfo{volume}{60}}, \bibinfo{pages}{741--744}
  (\bibinfo{year}{1976}).

\bibitem{mi2005}
\bibinfo{author}{B.~Mi}, \bibinfo{author}{J.~E. Michaels}, and
  \bibinfo{author}{T.~E. Michaels}, \enquote{\bibinfo{title}{An ultrasonic
  method for dynamic monitoring of fatigue crack initiation and growth}},
  \bibinfo{journal}{J. Acoust. Soc. Am.} \textbf{\bibinfo{volume}{119}},
  \bibinfo{pages}{74--85} (\bibinfo{year}{2001}).

\bibitem{wu1998}
\bibinfo{author}{T.~T. Wu} and \bibinfo{author}{T.~F. Lin},
  \enquote{\bibinfo{title}{The stress effect on the ultrasonic velocity
  variations of concrete under repeated loading}}, \bibinfo{journal}{ACI Mat.
  J.} \textbf{\bibinfo{volume}{95}}, \bibinfo{pages}{519--524}
  (\bibinfo{year}{1998}).

\bibitem{gret2006}
\bibinfo{author}{A.~Gret}, \bibinfo{author}{R.~Snieder}, and
  \bibinfo{author}{J.~Scales}, \enquote{\bibinfo{title}{Time-lapse monitoring
  of rock properties with coda wave interferometry}}, \bibinfo{journal}{J.
  Geophys. Res.} \textbf{\bibinfo{volume}{111}} (\bibinfo{year}{2006}).

\bibitem{payan:phd}
\bibinfo{author}{C.~Payan},
  \enquote{\bibinfo{title}{Non destructive evaluation of concrete. Potential of non-linear acoustics}}, \bibinfo{journal}{Ph. D. Thesis},
 \bibinfo{school}{Universit\'e de la M\'edit\'eran\'ee, Aix-En-Provence} (\bibinfo{year}{2007}).


\bibitem{weaver1982}
\bibinfo{author}{R.~L. Weaver}, \enquote{\bibinfo{title}{On diffuse waves in
  solid media}}, \bibinfo{journal}{J. Acoust. Soc. Am.}
  \textbf{\bibinfo{volume}{71}}, \bibinfo{pages}{1608--1609}
  (\bibinfo{year}{1982}).

\bibitem{lobkis2003}
\bibinfo{author}{O.~I. Lobkis} and \bibinfo{author}{R.~L. Weaver},
  \enquote{\bibinfo{title}{Coda-wave interferometry in finite solids: recovery
  of p-to-s conversion rates in an elastodynamic billiard}},
  \bibinfo{journal}{Phys. Rev. Lett.} \textbf{\bibinfo{volume}{90}},
  \bibinfo{pages}{254302} (\bibinfo{year}{2003}).

\bibitem{sens-schoenfelder2006}
\bibinfo{author}{C.~Sens-{Sch\"onfelder}} and \bibinfo{author}{U.~C. Wegler},
  \enquote{\bibinfo{title}{Passive image interferometry and seasonal variations
  of seismic velocities at merapi volcano, indonesia.}},
  \bibinfo{journal}{Geophys. Res. Lett.} \textbf{\bibinfo{volume}{33}},
  \bibinfo{pages}{L21302} (\bibinfo{year}{2006}).

\bibitem{note1}
The interpolation of the data at times $t(1-\nu)$ uses a spline interpolation algorithm. It should be noted that this procedure performs well even with our 4~MHz sampling rate. In fact, through numerical tests, we have found that the numerical precision of our processing procedure depends mainly on the signal-to-noise of the data and only feebly on the sampling rate. However even the effect of signal-to-noise is small; in the case of 1\% of electronic noise the error of the estimated relative velocity change $\nu_k$ is found to be much smaller than $10^{-6}$.

\bibitem{hadziioannou2009}
\bibinfo{author}{C.~hadziioannou}, \bibinfo{author}{E.~Larose},
  \bibinfo{author}{P.~Roux}, \bibinfo{author}{M.~Campillo},
  \enquote{\bibinfo{title}{Stability of Monitoring Weak Changes in Multiply Scattering Media with Ambient Noise Correlation: Laboratory Experiments.}}, \bibinfo{journal}{in preparation}
  

\bibitem{larose2006}
\bibinfo{author}{E.~Larose}, \bibinfo{author}{J.~De Rosny}, \bibinfo{author}{L.~Margerin}, \bibinfo{author}{D.~Anache}, \bibinfo{author}{P.~Gou\'edard},
\bibinfo{author}{M.~Campillo}, and \bibinfo{author}{B.~Van Tiggelen},
  \enquote{\bibinfo{title}{ Observation of multiple scattering of kHz vibrations in a concrete structure and application to monitoring weak changes}}, \bibinfo{journal}{Phys. Rev. E}
  \textbf{\bibinfo{volume}{73}}, \bibinfo{pages}{016609}
  (\bibinfo{year}{2006}).

\bibitem{leroy2008}
\bibinfo{author}{V.~Leroy} and \bibinfo{author}{A.~Derode},
  \enquote{\bibinfo{title}{Temperature-dependent diffusing acoustic wave
  spectroscopy with resonant scatterers}}, \bibinfo{journal}{Phys. Rev. E}
  \textbf{\bibinfo{volume}{77}}, \bibinfo{pages}{036602}
  (\bibinfo{year}{2008}).

\bibitem{yamamura2003}
\bibinfo{author}{K.~Yamamura}, \bibinfo{author}{O.~Sano},
  \bibinfo{author}{H.~Utada}, \bibinfo{author}{Y.~Takei},
  \bibinfo{author}{S.~Nakao}, and \bibinfo{author}{Y.~Fukao},
  \enquote{\bibinfo{title}{Long-term observation of in situ seismic velocity
  and attenuation}}, \bibinfo{journal}{J. Geophys. Res.}
  \textbf{\bibinfo{volume}{108}}, \bibinfo{pages}{2317} (\bibinfo{year}{2003}).






\bibitem{niu2008}
\bibinfo{author}{F.~Niu}, \bibinfo{author}{P.~G.~Silver},
  \bibinfo{author}{T.~M.~Daley}, \bibinfo{author}{X.~Cheng}, and
  \bibinfo{author}{E.~L.~Majer},
  \enquote{\bibinfo{title}{Preseismic velocity changes observed from active source monitoring
	at the Parkfield SAFOD drill site}}, \bibinfo{journal}{Nature}
  \textbf{\bibinfo{volume}{454}}, \bibinfo{pages}{204} (\bibinfo{year}{2008}).




\bibitem{brenguier2008b}
\bibinfo{author}{F.~Brenguier}, \bibinfo{author}{M.~Campillo},
  \bibinfo{author}{C.~Hadziioannou}, \bibinfo{author}{N.~M.~Shapiro},
  \bibinfo{author}{R.~M.~Nadeau}, and \bibinfo{author}{E.~Larose},
  \enquote{\bibinfo{title}{Postseismic relaxation along the San Andreas fault at Parkfield
	from continuous seismological observations}}, \bibinfo{journal}{Science}
  \textbf{\bibinfo{volume}{321}}, \bibinfo{pages}{1478-1481} (\bibinfo{year}{2008}).


\end{thebibliography}

\end{document}